\documentclass[aps,longbibliography,showpacs,twocolumn,superscriptaddress]{revtex4-1}
\usepackage{amsmath,amssymb,amsfonts,bm}
\usepackage{graphicx}
\usepackage{epstopdf}
\usepackage{dcolumn}
\usepackage{mathrsfs}
\usepackage{bbold}
\usepackage{dsfont}
\usepackage{float}
\usepackage[colorlinks=true,linkcolor=blue,citecolor=green, urlcolor=blue,bookmarks=false]{hyperref}
\usepackage{tgtermes}

\usepackage{changes}
\colorlet{Changes@Color}{red}

\begin{document}

\title{Weyl nodes with higher-order topology in an optically driven nodal-line semimetal}
\date{\today }
\author{Xiu-Li Du}
\altaffiliation{X.-L. D. and R. C. contributed equally to this work.}
\affiliation{Department of Physics, Hubei University, Wuhan 430062, China}
\author{Rui Chen}
\altaffiliation{X.-L. D. and R. C. contributed equally to this work.}
\affiliation{Shenzhen Institute for Quantum Science and Engineering and Department of Physics, Southern University of Science and Technology (SUSTech), Shenzhen 518055, China}
\author{Rui Wang}
\email[]{rcwang@cqu.edu.cn}
\affiliation{Institute for Structure and Function and Department of Physics, Chongqing University, Chongqing 400044, China}
\affiliation{Chongqing Key Laboratory for Strongly Coupled Physics, Chongqing 400044, China}
\affiliation{Center for Quantum Materials and Devices, Chongqing University, Chongqing 400044, China}
\author{Dong-Hui Xu}
\email[]{donghuixu@hubu.edu.cn}
\affiliation{Department of Physics, Hubei University, Wuhan 430062, China}

\begin{abstract}
Creating and manipulating topological states is a key goal of condensed matter physics. Periodic driving offers a powerful method to manipulate electronic states, and even to create topological states in solids. Here, we investigate the tunable Floquet states in a periodically driven higher-order nodal line semimetal with both spatial inversion and time-reversal symmetries. We found that the Floquet Weyl semimetal states, which support both one-dimensional hinge Fermi arc and two-dimensional surface Fermi arc states, can be induced in the higher-order nodal-line semimetal by shining circularly polarized light. Moreover, we show that the location of Weyl nodes and the curvature of surface Fermi arcs can be tuned by adjusting the propagation direction and incident angle of light.
\end{abstract}

\maketitle

\emph{\color{magenta}Introduction.}---Recent developments of topological insulators and topological superconductors have attracted
intensive interest in topological phases of matter. Creation and manipulation of topological states has become the central focus of condensed matter physics. Periodic driving through light fields opens a route towards engineering exotic Floquet topological states with highly tunability in solids via symmetry breaking or the modification of Dirac mass ~\cite{pssr.201206451,annurev-conmatphys-031218-013423}. The last decade has witnessed great progress made in the exploration of various Floquet topological states in optically driven systems \cite{PhysRevB.79.081406,PhysRevLett.105.017401,PhysRevB.82.235114,PhysRevLett.107.216601,PhysRevB.84.235108,PhysRevB.88.245422,PhysRevLett.113.266801,PhysRevB.90.115423,lindner2011floquet,PhysRevA.82.033429,PhysRevB.87.235131,PhysRevLett.106.220402,PhysRevX.3.031005,PhysRevLett.110.026603,PhysRevLett.110.200403,PhysRevLett.112.156801,Wang_2014,hubener2017creating,PhysRevB.94.041409,PhysRevLett.117.087402,PhysRevB.94.155206,PhysRevB.94.121106,PhysRevB.97.155152,PhysRevB.102.201105,PhysRevB.94.165436,PhysRevB.95.201115,PhysRevB.100.165302,PhysRevB.104.L081411}. For instance, circularly polarized light (CPL) has been predicted to gap out the Dirac cone and create the quantum anomalous Hall insulator in graphene by breaking time-reversal symmetry~(TRS)~\cite{PhysRevB.79.081406,PhysRevLett.105.017401,PhysRevB.82.235114,PhysRevLett.107.216601,PhysRevB.84.235108,PhysRevB.88.245422,PhysRevLett.113.266801,PhysRevB.90.115423}, while linearly polarized light can drive a band inversion and realize the Floquet topological insulator in semiconductor quantum wells~\cite{lindner2011floquet}. In a recent experiment, the light-induced anomalous Hall effect in graphene driven by a femtosecond pulse of CPL has been experimentally observed~\cite{mciver2020light}. Moreover, it has been reported experimentally that CPL can gap out the helical Dirac cones
of three-dimensional (3D) topological insulators and produce Floquet-Bloch states ~\cite{Wang453,mahmood2016selective}, which indicates the Floquet quantum anomalous Hall effect occurred in topological insulators. Apart from gapped Floquet topological phases, it has shown that Floquet Weyl semimetals (WSMs) can be generated in 3D topological insulators, Dirac semimetals, and nodal-line semimetals (NLSMs) by light driving \cite{Wang_2014,hubener2017creating,PhysRevB.94.041409,PhysRevLett.117.087402,PhysRevB.94.155206,PhysRevB.94.121106,PhysRevB.97.155152,PhysRevB.102.201105}.

Recently, higher-order topological insulators~\cite{Zhang2013PRL,Benalcazar2017Science,Langbehn2017PRL,Song2017PRL,Schindler2018SA,PhysRevLett.124.036803} with a brand-new bulk-boundary correspondence have become an exciting avenue
in the study of topological phases of matter. Compared with conventional first-order topological insulators, the dimensionality of topologically gapless boundary states in higher-order topological insulators is more than one dimension below that of the bulk.
Subsequently, the notion of higher-order topology was extended
to gapless semimetallic systems ~\cite{Ezawa2018PRL,Ezawa2018PRB2,Mao2018PRB,PhysRevResearch.1.032048,wieder2020strong,PhysRevLett.123.186401,PRL.125.126403,PhysRevLett.125.146401,PhysRevLett.125.266804,PhysRevLett.126.196402,rui2021intertwined}. Higher-order topological semimetals with point or line nodes exhibit unique boundary states
absent in conventional first-order topological semimetals, such as 1D gapless hinge Fermi arc states connect the projections of bulk nodes on the hinges~\cite{Mao2018PRB}. Moreover, the exploration of Floquet states with higher-order topology in periodically driven systems has attracted great interest recently \cite{PhysRevB.99.045441,PhysRevB.100.085138,PhysRevResearch.1.032045,PhysRevB.100.115403,PhysRevResearch.1.032045,PhysRevLett.123.016806,PhysRevLett.124.057001,PhysRevB.101.235403,PhysRevResearch.2.013124,PhysRevResearch.2.033495,PhysRevLett.124.216601,PhysRevB.103.L041402,PhysRevB.103.L121115,PhysRevB.103.115308,PhysRevB.104.L020302,PhysRevResearch.3.L032026,PhysRevB.103.184510}. So far, most attempts have focused on engineering a specific kind of Floquet higher-order topological state in periodically driven systems, but the investigation of light-driven systems with multiple higher-order topological states has been largely unexplored. Now, a natural question arises: Could we change the shape of bulk nodes with higher-order topology, such as from line nodes to point nodes, by an external periodic driving, such as CPL?

In this work, we study emergent Floquet topological states in a higher-order NLSM with both spatial inversion and time-reversal symmetries under CPL illumination. Such a higher-order NLSM has two nodal rings in its bulk, and supports boundary states, including 1D hinge Fermi arc states and 2D drumhead surface states. Remarkably, when shining CPL propagates along the $z$ axis, each nodal ring evolves into a pair of Weyl nodes located along the $k_z$ axis, and a Floquet higher-order WSM state emerges due to TRS breaking. The higher-order topology of the parent NLSM is inherited by the Floquet WSM state, resulting in 1D hinge Fermi arcs connecting the projections of two Weyl nodes from different pairs. Moreover, the Floquet WSM state hosts 2D surface Fermi arc states as well. When the propagation direction of light changes to the $x$ axis or the $y$ axis, a Floquet WSM state is also formed. In this case, the location of Weyl nodes deviates from the $k_z$ axis accordingly. Importantly, the location of Weyl nodes can change with tuning the incident angle of CPL, implying that we can control Weyl nodes by adjusting the propagation direction and incident angle of CPL.

\emph{\color{magenta}Model and method.}---To study the periodically driven higher-order NLSMs by CPL, we focus on a class of higher-order NLSMs with both spatial inversion and time-reversal symmetries~\cite{PhysRevLett.121.106403,PhysRevLett.123.186401,PRL.125.126403,lee2020two,PhysRevLett.126.196402}. Particularly, we consider a spinless four-band tight-binding model on the 3D cubic lattice, which can be treated as a stack of alternating layers. The Hamiltonian in momentum space reads~\cite{PRL.125.126403}
\begin{equation}\label{eq.1}
\begin{split}
H(\mathbf{k}) &= t\sin k_{x}\Gamma_{1}+t\sin k_{y}\Gamma_{2}\\
&+[ M-t(\cos k_{x}+\cos k_{y}+\cos k_{z})] \Gamma_{3} +H_{\text{mass}},
\end{split}
\end{equation}
where the Dirac matrices are defined as $\Gamma_{1}=\sigma_{0}\otimes\rho_{3}$, $\Gamma_{2}=\sigma_{2}\otimes\rho_{2}$, $\Gamma_{3}=\sigma_{0}\otimes\rho_{1}$, $\Gamma_{4}=\sigma_{1}\otimes\rho_{2}$, with $\sigma_{\mu}$ and $\rho_{\mu} ~(\mu=1,2,3)$ two sets of Pauli matrices and $\sigma_{0}$ the identity matrix. $\rho$ acts on the subspace formed by the two alternating layers, and $\sigma$ means the sublattice degrees of freedom of each layer. $t$ describes the hopping amplitude, and $M$ is the Dirac mass that controls the first-order topology. For the sake of simplicity, we set $t=1$ in the following calculations. The fourth term $H_{\text{mass}}=im(\Gamma_{1}\Gamma_{4}+\Gamma_{2}\Gamma_{4})$ is represented as a real matrix, standing for an additional mass that is responsible to generate an NLSM with higher-order topology. 

The inversion symmetry operator is $I=\Gamma_3$, and TRS operator is $T= \Gamma_{3}\mathcal{K}$ with $\mathcal{K}$ being the complex conjugate. In addition, the Hamiltonian also has chiral symmetry $C=\Gamma_{4}$ and mirror symmetry $M_{x\bar{y}}=\Gamma_{1} \Gamma_{4}+\Gamma_{2} \Gamma_{4}$ under which the Bloch Hamiltonian obeys: $M_{x\bar{y}}H(\mathbf{k})M_{x\bar{y}}^{-1}=H (-k_y,-k_x, k_z)$. 

In the absence of $H_{\text{mass}}$, the Hamiltonian describes a 3D first-order Dirac semimetal, which has two Dirac nodes residing at $k_z=\pm k_{z}^0=\pm \arccos(M-2)$.
The low-energy effective Hamiltonians around the Dirac points are given by
\begin{equation}\label{DSM}
	H_{\text{DSM}}(\mathbf{q})=q_{x}\Gamma_{1}+q_{y}\Gamma_{2}\pm\sqrt{(3-M)(M-1)}q_{z}\Gamma_{3},
\end{equation}
where $\mathbf{q}$ is momentum measured relative to the Dirac points. After turning $H_{\text{mass}}$ on, the Dirac semimetal evolves into a higher-order NLSM with two nodal rings on the mirror invariant plane.
\begin{figure}[ptb]
	\centering
	\includegraphics[width=8cm]{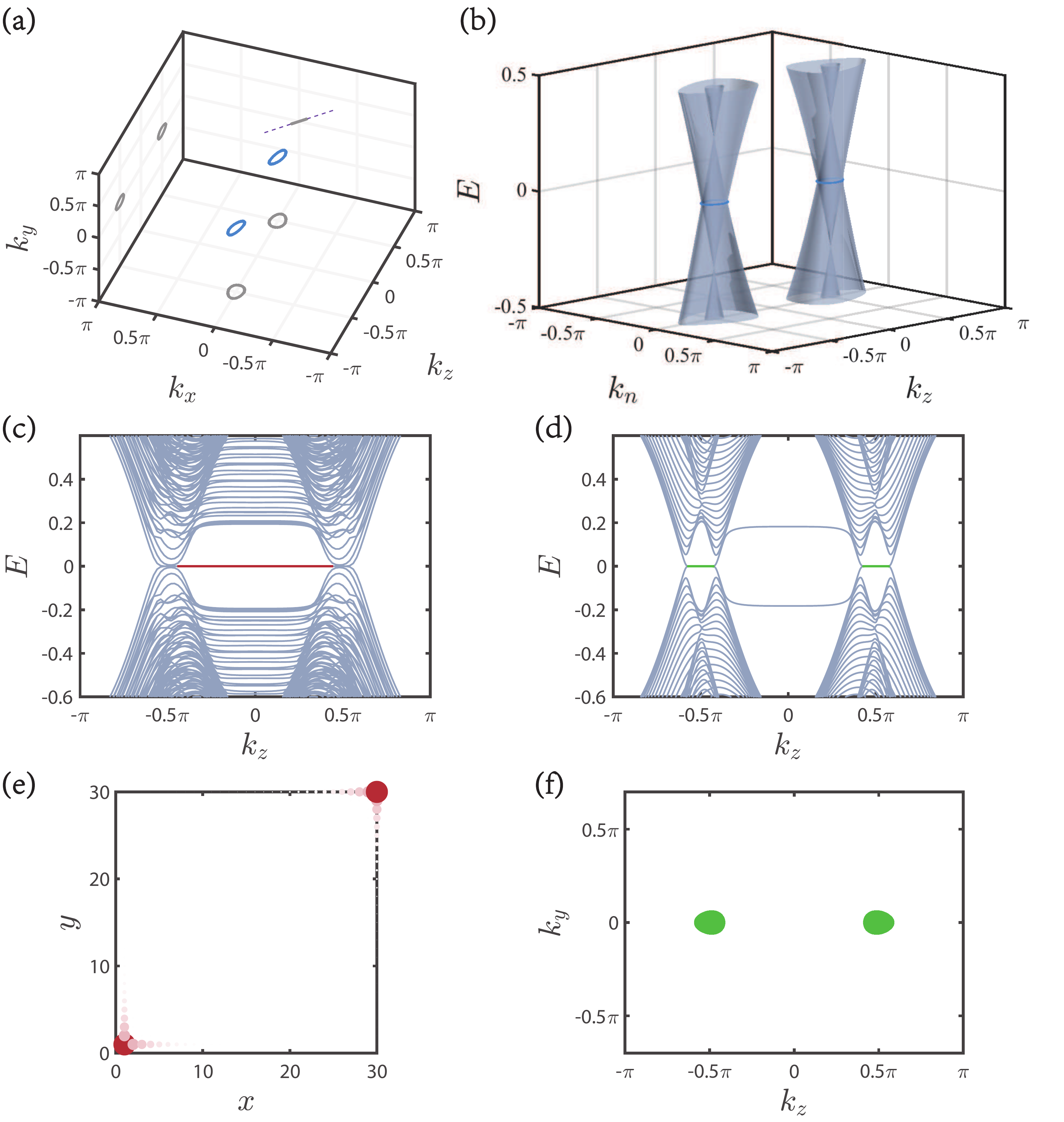}
	\caption{Electronic structure of the undriven NLSM obtained by the tight-binding model described by Eq.~\ref{eq.1}. (a) Two higher-order nodal rings (blue) in the 3D Brillouin zone. The gray rings are the projections of the nodal rings on $k_x$-$k_y$, $k_x$-$k_z$ and $k_y$-$k_z$ planes, respectively. (b) The bulk energy spectrum as functions on the $k_{\mathbf{n}}$-$k_z$ plane with $\mathbf{k}_\mathbf{n}$ along the $k_x=-k_y$ direction. (c) The energy spectrum as a function of $k_{z}$ with the open boundary conditions along both the $x$ and $y$ directions. The red solid lines represent the hinge Fermi arc states. (d) The energy spectrum as a function of $k_{z}$ with the open boundary condition along the $x$ direction for $k_y=0$. The green lines represent the drumhead surface states. (e) The probability distribution of the hinge Fermi arc states of $k_{z}=0$. (f) The spectral density, calculated by the surface Green's function method, in the surface Brillouin zone defined on $k_y$-$k_z$ plane for $E=0$. The drumhead surface states appear in the regions bounded by the two projected nodal rings. The mass parameters are set as $M=2$, $m=0.2$.}
	\label{fig_withoutlight}
\end{figure}

Before studying the Floquet states induced by periodic driving, it is helpful to discuss the band structure of the undriven higher-order NLSM described by Eq. \ref{eq.1}. Figures \ref{fig_withoutlight}(a) and \ref{fig_withoutlight}(b) show the two bulk nodal rings of the NLSM lying on the mirror plane $k_x$=$-k_y$ plane of 3D Brillouin zone and the band structure around them. Figure \ref{fig_withoutlight}(c) plots the energy spectrum along the $k_z$ direction for the system with open boundaries in both the $x$ and $y$ directions. A pair of hinge Fermi arc states with zero-energy flat bands (marked in red) terminate on the projection of nodal rings, which are the hallmark feature of higher-order topology in NLSMs. As shown in Fig~\ref{fig_withoutlight}(e), the hinge Fermi arc states are located on two mirror-symmetric off-diagonal hinges. Meanwhile, this kind of higher-order NLSM also supports flatband drumhead surface states in the region enclosed by the projections of nodal rings onto the surface Brillouin zone, as displayed in Fig.~\ref{fig_withoutlight}(f). With open boundary in the $x$ direction, the spectrum along the $k_z$ direction for $k_y=0$ is depicted in Fig.~\ref{fig_withoutlight}(d). The eigenvalues of drumhead surface states (marked in green) are pinned to zero energy due to chiral symmetry.

The circularly polarized driving field can be expressed as $\mathbf{E}(\tau)=-\partial_\tau \mathbf{A}(\tau)$, where $\mathbf{A}(\tau)=\mathbf{A}(\tau+T)$ is a time-periodic, spatially homogeneous vector potential of period $T=2\pi/\omega$, and $\omega$ is the frequency of light. The coupling to driving field $\mathbf{E}(\tau)$ enables electrons to acquire a phase factor as they hop between different lattice sites. Specifically, the phase factor can be obtained by the Peierls substitution: $t\rightarrow t\exp[-i\int\nolimits_{\mathbf{r}_{j}}^{\mathbf{r}_{k}}\mathbf{A}(\tau)\cdot d\mathbf{r}]$, where $\mathbf{r}_j$ is the coordinate of lattice site $j$. In this paper, we use the natural units $e=\hbar=c=1$. In the presence of $E(\tau)$, the Hamiltonian of the driven NLSMs becomes periodic and fulfills $H(\tau)=H(\tau+T)$.

 According to the Floquet theory~\cite{PhysRev.138.B979,PhysRevA.7.2203}, there exits a set of solutions of the time-dependent Schr\"{o}dinger equation $|\Psi(\tau) \rangle =e^{-i\epsilon \tau}|\Phi(\tau)\rangle$, where $\epsilon$ is the Floquet quasienergy and the time-periodic function $|\Phi(\tau)\rangle=|\Phi(\tau+T)\rangle$ is called the Floquet state. Using the Fourier transformation $H(\tau)=\sum_n e^{-in\omega\tau} H_n$ and $|\Phi(\tau)\rangle=\sum_n e^{-in\omega \tau}|\Phi^n\rangle$, the time-dependent Schr\"{o}dinger equation is mapped to a static eigenvalues problem
\begin{equation}
	\sum_m(H_{n-m}-m\omega \delta_{mn})|\Phi_\alpha^m\rangle=\epsilon_\alpha|\Phi_\alpha^m\rangle,
\end{equation}
in the extended Floquet (or Sambe) space. Throughout, we focus on the situation where electron transitions through the absorption or emission of photons are very unlikely. This situation occurs for high frequency light. We can arrive at an effective Hamiltonian in the high frequency limit~\cite{bukov2015universal,Eckardt_2015}
 \begin{equation}\label{eq.4}
	H_{\text{eff}}=H_{0}+\sum_{n\neq 0}\frac{\left[ H_{-n},H_{n}\right] }{2n\omega }+\mathcal O(\omega ^{-2}).
\end{equation}
In numerical calculations, we can choose the proper maximum value of $n$ by checking whether the results are converged.

\emph{\color{magenta}Light propagating along the $z$ direction.}---
First, we introduce the CPL along the $z$-axis $\mathbf{A}=A \left( \eta \sin \omega \tau,\cos \omega \tau,0\right)$ with $\eta=\pm$ labeling the handedness. Substituting Eq.~\ref{eq.1} into Eq.~\ref{eq.4}, we get an effective Hamiltonian of the driven higher-order NLSM in the high frequency limit as follows
\begin{eqnarray} \label{effectiveHz}
		H_{\text{eff}}(\mathbf{k})\!&=&t\mathcal{J}_{0}(A)\sin k_{x}\Gamma_{1}+t\mathcal{J}_{0}(A)\sin k_{y}\Gamma_{2}+H_{\text{mass}}\\
		&+&\big( M-t\mathcal{J}_{0}(A)(\cos k_{x}+\cos k_{y})-t\cos k_{z}\big) \Gamma_{3}\nonumber \\
		&\!+\!&\sum_{n\in \text{odd}, n>0}\frac{2i\eta t^2\mathcal J_{n}^{2}(A)}{n \omega }\sin\frac{n\pi }{2}\Big(\cos k_{x} \cos k_{y}[\Gamma_{1},\Gamma_{2}]\nonumber \\
		&+&\cos k_{x} \sin k_{y}[\Gamma_{1},\Gamma_{3}]+\sin k_{x} \cos k_{y}[\Gamma_{3},\Gamma_{2}]\Big),\nonumber
\end{eqnarray}
where $\mathcal{J}_{n}(A)$ is the $n$-th Bessel function of the first kind. Comparing the effective Hamiltonian with its undriven case, we can see that the CPL not only renormalizes the original hoppings along the $x$ and $y$ directions, but also generates correction terms breaking TRS and chiral symmetry. However, the product of chiral symmetry and TRS $CT$ as well as mirror symmetry $M_{x\bar{y}}$ are still preserved.

\begin{figure}[ptb]
	\centering
	\includegraphics[width=8cm]{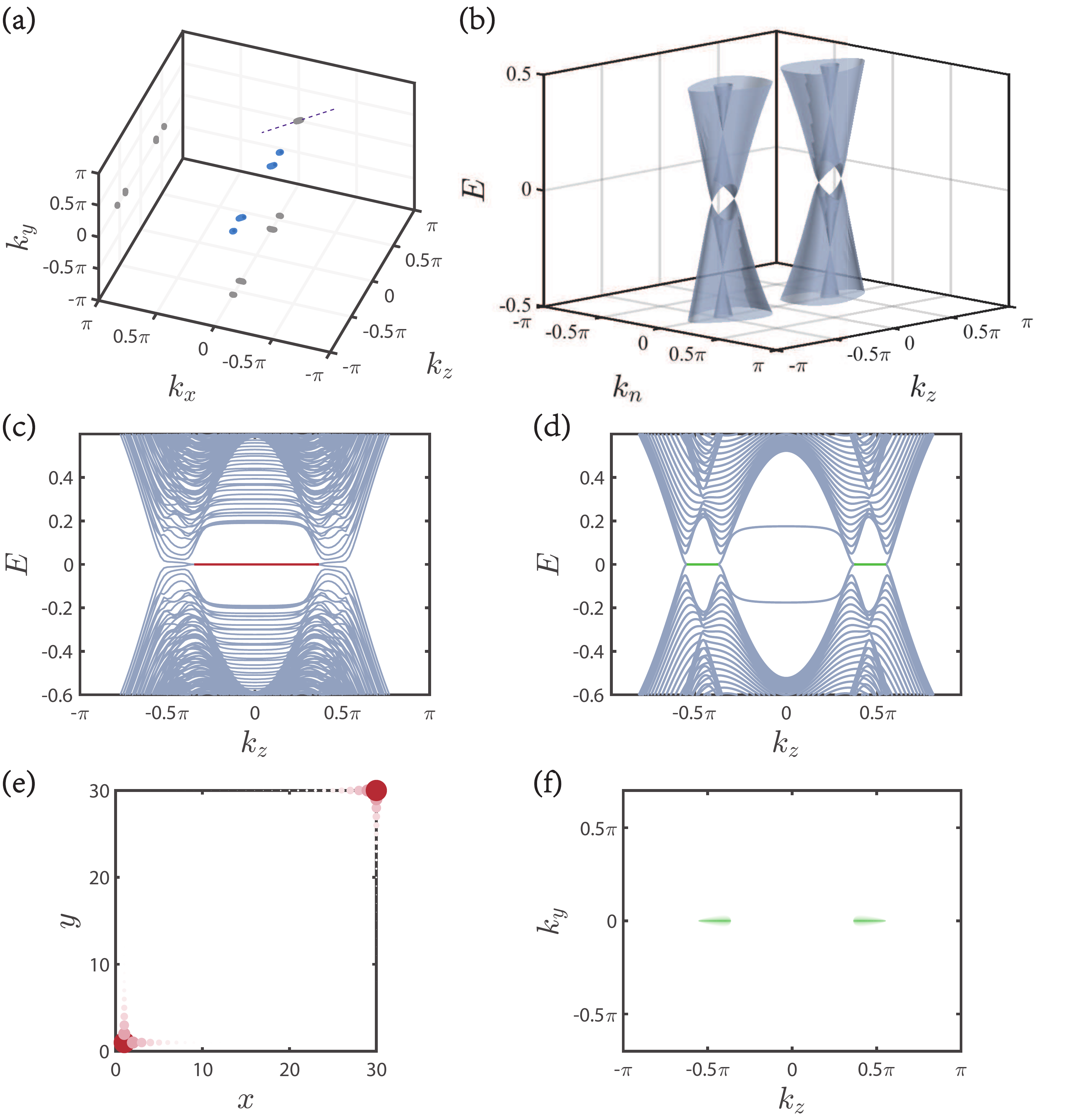}
	\caption{Higher-order NLSM under CPL propagates along the $z$ axis. (a) The light-induced Weyl points (blue dots) in the 3D Brillouin zone. The gray dots are the projections of the Weyl points on $k_x$-$k_y$, $k_x$-$k_z$ and $k_y$-$k_z$ planes, respectively. (b) The bulk energy spectrum as functions on the $k_{\mathbf{n}}$-$k_z$ plane with $\mathbf{k}_\mathbf{n}$ along the $k_x=-k_y$ direction. (c) The energy spectrum as a function of $k_{z}$ with the open boundary conditions along the $x$ and $y$ directions. The red solid lines represent the hinge Fermi arc states. (d) The energy spectrum as a function of $k_{z}$ with the open boundary condition along the $x$ direction and $k_y=0$. The Green lines are surface Fermi arc states. (e) The probability distribution of the hinge Fermi arc states at $k_{z}=0$. (f) The location of surface Fermi arc states terminated at the projection of Weyl points on the surface Brillouin zone defined on the $k_y$-$k_z$ plane. The two line segments marked by green lines show the surface spectral density for $E=0$, which is obtained by the surface Green's function method. We choose parameters as $M=2$, $m=0.2$, $A=0.5$ and $\omega=3$.}
	\label{fig_light z}
\end{figure}

Figure~\ref{fig_light z} shows the band structure obtained by the effective Hamiltonian Eq.~\ref{effectiveHz}. Interestingly, each bulk nodal ring evolves into a pair of Weyl nodes residing along the mirror invariant line $k_x=k_y=0$, depicted in Figs.~\ref{fig_light z}(a) and~\ref{fig_light z}(b). The Weyl nodes inhibit the higher-order topology from the nodal rings, resulting in a Floquet higher-order WSM state. As is evident in Figs.~\ref{fig_light z}(c) and~\ref{fig_light z}(e), the Floquet higher-order WSM state hosts 1D hinge Fermi arcs with flat bands connecting by the projection of two middle Weyl nodes from two different pairs. For each slice with a given $k_z$ within the two middle Weyl points, the effective Hamiltonian Eq.~\ref{effectiveHz} reduces to a 2D second-order topological insulator with mirror invariant off-diagonal corner states. Moreover, such a light-induced higher-order WSM state also exhibits 2D surface Fermi arc states with zero energy dispersionless flat bands between a pair of projected Weyl nodes [see Fig.~\ref{fig_light z}(d)] due to the existence of spectral symmetry $CT$ and mirror symmetry $M_{x\bar{y}}$. Accordingly, the surface Fermi arcs lose their curvature and form straight line segments [see Fig.~\ref{fig_light z}(f)] which end at the projections of pairs of bulk Weyl nodes along the $k_y=0$ line in the surface Brillouin zone of the $k_y$-$k_z$ plane.

To better understand the Floquet WSM sates, we appeal to the low-energy continuum model of the driven NLSM within the framework of Floquet theory. The periodic driving field is introduced in the Hamiltonian by using the minimal coupling substitution $\mathbf{q} \rightarrow\mathbf{q}-\mathbf{A}(\tau)$. In the high frequency regime, the effective Floquet Hamiltonian up to the leading order is expressed as ~\cite{annurev-conmatphys-031218-013423} $H_\text{F}=H_{0}+\frac{[H_{-1},H_{1}]}{ \omega }$.
Substituting the continuum model $H_{\text{c}}(\mathbf{q})=H_{\text{DSM}}(\mathbf{q})+H_{\text{mass}}$ into $H_F$, we obtain the correction induced by CPL: $ \Delta H=-\frac{i \eta A^{2}}{2\omega}[\Gamma_{1},\Gamma_{2}].$
 $\Delta H$ is independent of $\mathbf{q}$ as $H_{\text{DSM}}(\mathbf{q})$ is only expanded to linear order in $\mathbf{q}$. The correction $\Delta H$ enables each nodal ring to evolve into a pair of Weyl nodes located at
\begin{equation}
q_x^0=q_y^0=0,\;\;\;q_{z}^0=\pm \sqrt{\frac{\eta^{2}A^{4}+2\omega^{2}m^{2}}{\omega^{2}(3-M)(M-1)}}.
\end{equation}
This result is consistent with that obtained by using the tight-binding lattice model.
\begin{figure*}[ptb]
	\centering
	\includegraphics[width=16cm]{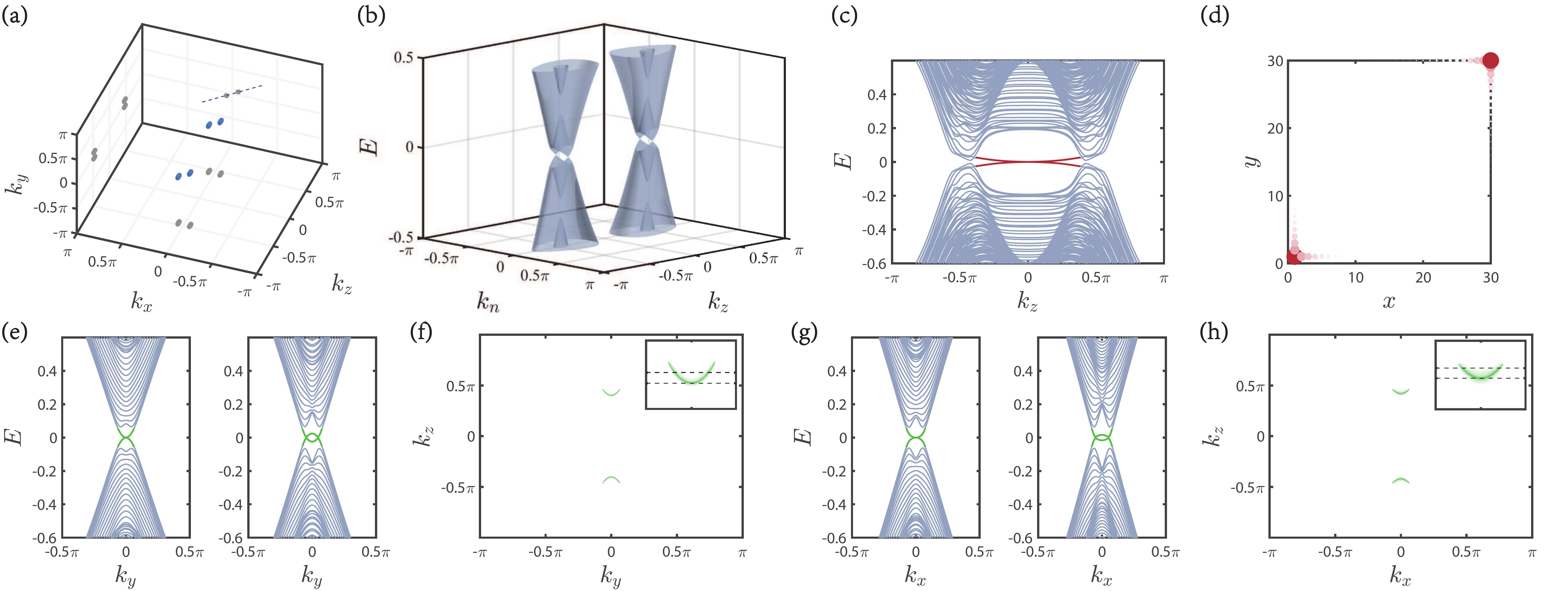}
	\caption{Band structure of the driven higher-order NLSM under CPL propagates along the $x$-axis. (a) The Floquet Weyl points (blue dots) in the 3D Brillouin zone. The gray dots are the projections of the two pairs of Weyl points on $k_x$-$k_y$, $k_x$-$k_z$ and $k_y$-$k_z$ planes, respectively. (b) The bulk energy spectrum around the $k_{\mathbf{n}}$-$k_z$ plane on which the Weyl nodes are located. Here $\mathbf{k}_\mathbf{n}$ is along $k_x=-0.991k_y$ direction, slightly deviating from the mirror plane. (c) The energy spectrum as a function of $k_{z}$ with the open boundary conditions along the $x$ and $y$ directions. The red solid lines mark the off-diagonal hinge Fermi arc states. (d) The probability distribution of the hinge Fermi arc states at $k_{z}=0$. (e) The energy spectra along the dashed line cuts shown in the inset of (f) for the open boundary condition along the $x$ direction. The surface Fermi arc states are marked by green color. (f) The location of surface Fermi arc states terminated at the projection of Weyl points on the surface Brillouin zone defined on the $k_y$-$k_z$ plane. The two curves marked in green show the surface spectral density for $E=0$, which is obtained by the surface Green's function method. The inset shows a zoomed-in view of the up Fermi arc. (g) The energy spectra along the dashed line cuts shown in the inset of (h) for the open boundary condition along the $y$-direction. The surface Fermi arc states are marked by green color. (h) The location of surface Fermi arcs on the $k_x$-$k_z$ plane. We choose parameters the same as that used in Fig.~\ref{fig_light z}.}
	\label{fig_light x}
\end{figure*}

\emph{\color{magenta}Light propagating along the $x$ direction.}---To see how the propagation direction of CPL affects the properties of Floquet states, we consider CPL along the $x$ axis $\mathbf{A}=A \left( 0,\eta \sin \omega \tau,\cos \omega \tau \right)$. In this case, the effective tight-binding Hamiltonian in the high frequency limit becomes
\begin{eqnarray}\label{effectiveHx}
	 H_\text{eff}(\mathbf{k})&\!=\!& t\sin k_{x}\Gamma_{1}+t\mathcal{J}_{0}(A)\sin k_{y}\Gamma_{2}+H_{\text{mass}}\\
		&+&\big( M-t\cos k_{x}-t\mathcal{J}_{0}(A)(\cos k_{y}+\cos k_{z})\big) \Gamma_{3}\nonumber \\
		&\!-\!&\sum_{n\in \text{odd}, n>0}\frac{2i\eta t^{2}\mathcal J_{n}^{2}(A)}{n \omega }\sin\frac{n\pi }{2} \cos k_{y} \sin k_{z}[\Gamma_{3},\Gamma_{2}].\nonumber 
\end{eqnarray}
The CPL along the $x$ axis renormalizes the hoppings in the $y$ and $z$ axes, meanwhile it also imposes a correction that breaks TRS and chiral symmetry to the undriven Hamiltonian.  Besides, mirror symmetry $M_{x\bar{y}}$ is also broken by the CPL in the $x$ direction. By diagonalizing the effective Hamiltonian, we plot the energy spectrum in Fig.~\ref{fig_light x}. Figures~\ref{fig_light x}(a) and~\ref{fig_light x}(b) show the location of two pairs of Weyl nodes in the 3D Brillouin zone and the bulk Weyl cone structure around the Weyl nodes, respectively. We found that the locations of two pairs of Weyl nodes deviate from the mirror plane $k_x=-k_y$ owing to mirror symmetry breaking. Again, the light-induced WSM state displays 1D off-diagonal hinge Fermi arcs, as shown in Figs.~\ref{fig_light x}(c) and \ref{fig_light x}(d), indicating the unconventional bulk-boundary correspondence originated from higher-order topology. Yet the hinge Fermi arc states exhibit dispersive bands in the current case. In addition, the Floquet WSM supports curved surface Fermi arcs [see Figs.~\ref{fig_light x}(f) and \ref{fig_light x}(h)], which is different from the case of Floquet WSM induced by the CPL along the $z$ axis where the surface Fermi arcs states show straight line segments. To better understand the formation of surface Fermi arcs, we show the spectra of surface Fermi arc states in Figs.~\ref{fig_light x}(e) and \ref{fig_light x}(g).

Similarly, on the basis of the continuum model, we obtain the correction
$\Delta H=\frac{i \eta A^{2}\sqrt{(3-M)(M-1)}}{2 \omega }[\Gamma_{3},\Gamma_{2}].$
 Thereby, the locations of Weyl nodes are at $\mathbf{q}=(q_x^0,q_y^0,0)$ with
\begin{eqnarray}
q_x^0&=&\pm \Big(4m^{4}+B^{2}\Big)^{\frac{1}{4}}\cos\Big[\frac{1}{2}\arg(-2im^{2}-B)\Big],\nonumber \\
q_y^0&=&\pm \Big(4m^{4}+B^{2}\Big)^{\frac{1}{4}}\sin\Big[\frac{1}{2}\arg(-2im^{2}-B)\Big],
\end{eqnarray}
where $B=A^{4}(3-4M+M^{2})/\omega^{2}$.

\emph{\color{magenta}Light propagating along the $y$ direction.}---Here, we consider the CPL along the $y$ axis:
$\mathbf{A}=A\left(\cos \omega \tau,0,\eta \sin\omega \tau\right) $. Correspondingly, we have the following effective Hamiltonian in the high frequency limit
\begin{eqnarray}
		H_\text{eff}(\mathbf{k})&=& t\mathcal{J}_{0}(A)\sin k_{x}\Gamma_{1}+t\sin k_{y}\Gamma_{2}+H_{\text{mass}}\\
		&+&\big( M-t\cos k_{y}-t\mathcal{J}_{0}(A)(\cos k_{x}+\cos k_{z})\big) \Gamma_{3}\nonumber \\
		&-&\sum_{n\in \text{odd}, n>0}\frac{2i\eta t^{2}\mathcal J_{n}^{2}(A)}{n \omega }\sin\frac{n\pi }{2} \cos k_{x} \sin k_{z}[\Gamma_{1},\Gamma_{3}].\nonumber 
\end{eqnarray}
Although this effective Hamiltonian looks different from that in Eq.~\ref{effectiveHx} for the CPL along the $x$ axis, the location of Weyl nodes and boundary states are quite similar to that are exhibited in Fig.~\ref{fig_light x}. The correction term obtained by the continuum model is $\Delta H=\frac{i  \eta A^{2}\sqrt{(3-M)(M-1)}}{2 \omega }[\Gamma_{1},\Gamma_{3}].$

Besides, we found that the higher-order NLSM is stable when CPL is incident normally on the mirror plane containing the two nodal rings.

\begin{figure}[hbp]
	\centering
	\includegraphics[width=6cm]{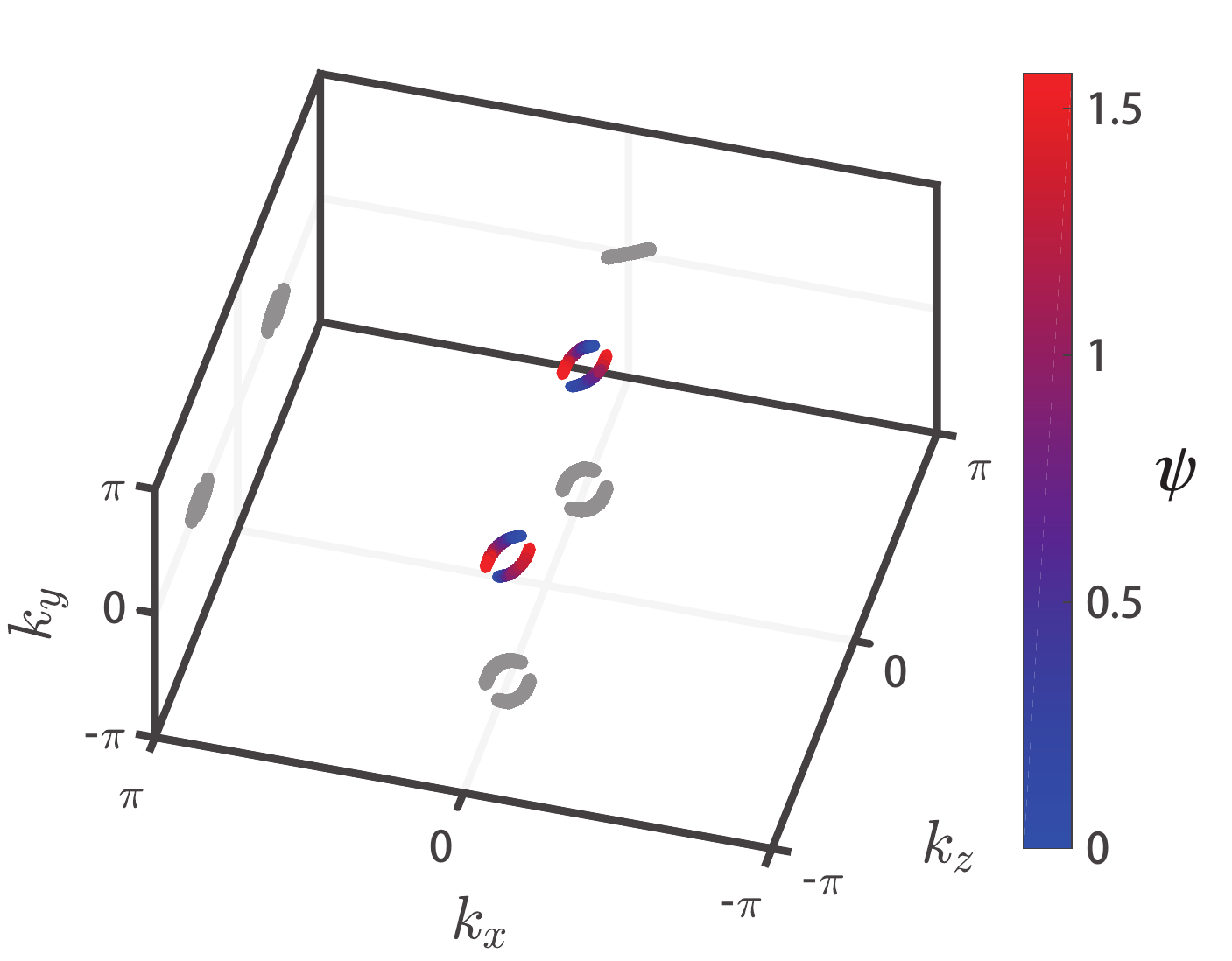}
	\caption{The position of the light-induced Weyl points~(colored dots) obtained by Eq.~\ref{Heffxz} versus the incident angle $\psi$ of CPL in the $x$-$z$ plane. The gray dots are the projections of the Weyl points on $k_x$-$k_y$, $k_x$-$k_z$ and $k_y$-$k_z$ planes, respectively. We choose model parameters the same as that used in Fig.~\ref{fig_light z}.}
	\label{fig_light_xz}
\end{figure}

\emph{\color{magenta}Light propagating on the $x$-$z$ plane.}---Furthermore, we can rotate the propagation direction of CPL to the $x$-$z$ plane, the vector potential can be expressed as $\mathbf{A}=\left( A_{x}\eta \sin \left( \omega \tau\right),A_{y}\cos \left(\omega \tau\right), A_{z}\eta \sin \left( \omega \tau\right)\right)$, where $A_{x}=A \cos \left( \psi\right)$,  $A_{y}=A$,  $A_{z}=A \sin \left( \psi\right)$ with $\psi$ the incident angle off the $z$ axis. When $\psi=0$ and $\psi=\pi/2$, it reduces to the case of CPL along the $z$ axis and the negative $x$ axis, respectively. The effective Floquet-Bloch Hamiltonian is given by
\begin{eqnarray}\label{Heffxz}
		&&H_{\text{eff}}(\mathbf{k})=t\mathcal{J}_{0}(A_{x})\sin k_{x}\Gamma_{1}+t\mathcal{J}_{0}(A_{y})\sin k_{y}\Gamma_{2}+H_{\text{mass}} \nonumber \\
		&+&\!\Big( M\!\!-\!t\big(\mathcal{J}_{0}(A_{x})\cos k_{x}\!-\!\mathcal{J}_{0}(A_{y})\cos k_{y}
		\!-\!\mathcal{J}_{0}(A_{z})\cos k_{z}\big)\Big) \Gamma_{3}\nonumber \\
		&+&\!\!\sum_{n\in \text{odd}, n>0} \!\! \frac{2i\eta t^2\mathcal J_{n}(A_{x})\mathcal J_{n}(A_{y})}{n \omega }\sin\frac{n\pi }{2}\nonumber \\
		&\times&\Big(\cos k_{x} \cos k_{y}[\Gamma_{1},\Gamma_{2}]+\cos k_{x} \sin k_{y}[\Gamma_{1},\Gamma_{3}]\nonumber \\
		&+&\sin k_{x} \cos k_{y}[\Gamma_{3},\Gamma_{2}]\Big)
		\!\!+\sum_{n\in \text{odd}, n>0}\!\! \frac{2i\eta t^2\mathcal J_{n}(A_{z})\mathcal J_{n}(A_{y})}{n \omega }\nonumber \\
		&\times&\sin\frac{n\pi }{2}\sin k_{z} \cos k_{y}[\Gamma_{3},\Gamma_{2}].
\end{eqnarray}

Figure \ref{fig_light_xz} maps out the evolution of Weyl nodes as the incident angle $\psi$ varies by using this effective Hamiltonian. The result suggests that we can manipulate the postilion of Weyl nodes by tuning the propagation direction of the incident CPL.

The leading order correction term to the continuum model can be expressed as
$\Delta H \!=\!\!-\!\frac{i  \eta A_{x}A_{y}}{2 \omega}[\Gamma_{1},\Gamma_{2}]\!-\!\frac{i \eta A_{y}A_{z}\sqrt{(3\!-\! M)(M \!-\! 1)}}{2 \omega }[\Gamma_{3},\Gamma_{2}].$

\emph{\color{magenta}Conclusions.}---In this study, we uncovered the emerging Floquet WSM states in the parity-time invariant higher-order NLSM under CPL by breaking TRS. The Floquet WSM states accommodate 1D off-diagonal gapless hinge Fermi arcs caused by inherited higher-order topology from the parent NLSM. Meanwhile, the Floquet WSM states also support 2D surface Fermi arcs as in conventional WSMs. We also show that the location of Weyl nodes can be tuned by adjusting the propagation direction and incident angle of CPL. 

We would like to emphasize that, unlike the Floquet WSMs in periodically driven ordinary NLSMs~\cite{PhysRevB.94.041409,PhysRevLett.117.087402,PhysRevB.94.155206,PhysRevB.94.121106,PhysRevB.97.155152,PhysRevB.102.201105}, the higher-order Floquet WSM here has the coexisting 1D hinge and 2D surface Fermi arc states. Furthermore, we unveiled the dispersion of Fermi arc states can be adjusted by tuning the propagation direction of CPL. Higher-order WSMs have so far been realized only in acoustic crystals~\cite{luo2021observation,wei2021higher}, while our work provides a feasible way to realize tunable higher-order WSM states in non-equilibrium electronic systems. Also, it serves as a reference for future experiments on periodically driven higher-order NLSMs, such as XTe$_2$ (X=Mo, W)~\cite{PhysRevLett.123.186401} and 3D ABC stacked graphdiyne~\cite{PhysRevLett.121.106403,lee2020two,PhysRevLett.128.026405}.

\emph{\color{magenta}Acknowledgments.}---The authors acknowledge the support by the NSFC (under Grant Nos. 12074108, 11704106, 11974062 and 12047564), the China Postdoctoral Science Foundation (Grant No. 2019M661678), the SUSTech Presidential Postdoctoral Fellowship, the Chongqing Natural Science Foundation (Grant No. cstc2019jcyj-msxmX0563), the Fundamental Research Funds for the Central Universities of China (Grant No. 2020CDJQY-A057), and the Beijing National Laboratory for Condensed Matter Physics.

\bibliography{bibfile}

\end{document}